\documentclass[twocolumn,pre,showpacs]{revtex4-1}
\usepackage{graphicx}
\usepackage{amsmath}
\usepackage{amsfonts}
\usepackage{mathtools}
\usepackage[utf8]{inputenc}

\begin{document}
\title{Stochastic thermodynamics with information reservoirs}

\author{Andre C. Barato and Udo Seifert}
\affiliation{ II. Institut f\"ur Theoretische Physik, Universit\"at Stuttgart, 70550 Stuttgart, Germany}

\parskip 1mm
\def\d{{\rm d}}
\def\Ps{{P_{\scriptscriptstyle \hspace{-0.3mm} s}}}
\def\MF{{\mbox{\tiny \rm \hspace{-0.3mm} MF}}}
\def\i{\text{\scriptsize $\cal{I}$}} 

\begin{abstract}
We generalize stochastic thermodynamics to include information reservoirs. Such information reservoirs,
which can be modeled as a sequence of bits, modify the second law. For example, work extraction from a system
in contact with a single heat bath becomes possible if the system also interacts with an information reservoir. 
We obtain an inequality, and the corresponding fluctuation theorem, generalizing the standard entropy production of stochastic thermodynamics. From this inequality 
we can derive an information processing entropy production, which gives the second law in the presence of information reservoirs.   
We also develop a systematic linear response theory for information processing machines. For an uni-cyclic machine powered by an information reservoir, 
the efficiency at maximum power can deviate from the standard value $1/2$. For the case where energy is consumed to erase the tape, 
the efficiency at maximum erasure rate is found to be $1/2$. 
\end{abstract}

\maketitle
\section{Introduction}

Including information processing into thermodynamics has received much attention since its starting point with Maxwell's demon \cite{leff03,maru09}.
The first considerations of ``violations'' of the second law induced by an external controller were restricted to thought experiments that could not be
reproduced in the laboratory. The situation has recently changed, as experiments with colloids allow the verification of Landauer's principle \cite{beru12} and
the conversion of information into work \cite{toya10a}, for example. Moreover, this fundamental generalization of thermodynamics should play an important role in improving our 
understanding of problems like computer dissipation \cite{benn82} and cellular sensing \cite{bial12}.

One approach to study the relation between information and thermodynamics is to consider  
feedback driven systems \cite{bech05}, for which a controller measures the state of the system and changes the protocol 
according to the measurement outcome and some probabilistic rule. The second law of thermodynamics for feedback driven systems also includes 
the mutual information between the system and controller \cite{cao09}. Prominently among the many recent works on 
the relation between information and thermodynamics \cite{touc00,touc04,alla08,alla09,saga10,ponm10,horo10,horo11,horo11a,gran11,espo11,abre11,abre11a,baue12,kund12,stil12,muna12,
kish12,saga12,saga12b,espo12b,stra13,stra13a,horo13,gran13,dian13a,saga13,ito13,saga14,hart14,baue14,horo14}, Sagawa and Ueda obtained a fluctuation relation for feedback driven systems generalizing this 
second law \cite{saga10}.

A different approach to study the thermodynamics of information processing has been recently proposed by Mandal and Jarzynski (MJ) \cite{mand12}. They introduced 
a simple model for a thermodynamic system interacting with a tape (a sequence of bits), where work can be extracted from a system in contact with a single 
heat bath by increasing the Shannon entropy of the tape, i.e., by writing information on the tape. Within the MJ model a tape full of zeros is a thermodynamic resource that can be consumed to do useful
work, an idea expressed by Bennett some time ago \cite{benn82}. Two generalizations of the MJ model feature a tape that can move in both directions \cite{bara13} and a thermal tape with non-zero temperature \cite{hopp14}.
Furthermore, a similar model for a refrigerator powered by writing information on a tape was introduced in \cite{mand13}. 

More generally, this tape can be viewed as an information reservoir \cite{deff13,bara14}, which is a reservoir that only changes the entropy balance. Thus it must be accounted for
in the second law while leaving the first law intact, as no energy is exchanged between the information reservoir and the system. Deffner and Jarzynski have obtained the generalizations 
of the second law with an information reservoir using an Hamiltonian framework \cite{deff13}. We have shown that the theory of stochastic thermodynamics could be
generalized to include an information reservoir \cite{bara14}. 

In this paper we further extend the result obtained in \cite{bara14}, by proving an inequality that allows us to generalize stochastic thermodynamics to the presence of several information reservoirs. 
This generalization is achieved by introducing an information processing entropy production (IP-entropy production), which takes into account information reservoirs interacting with the system. 
A master fluctuation theorem leading to our generalized inequality is also proved. Furthermore, we obtain the modified forms for the second and first law in the presence of information reservoirs and
demonstrate with specific examples that our formalism can be used to study a generic thermodynamic system interacting with information reservoirs. 

A precursor in analyzing thermodynamic systems out of equilibrium is linear response theory \cite{groot,pott09}. Whereas even fluctuation theorems are available 
for information processing machines \cite{saga10,horo10,abre11,saga12}, a systematic linear response theory has not yet been considered, apart from our case study in \cite{bara13}.
Our present framework allows for the development of such a linear response theory.
We obtain a general form for the IP-entropy production in terms of the affinities and the Onsager matrix. For uni-cyclic machines, we  
show that an IP-efficiency, involving information processing, at maximum power varies between $1/2$ and $2/3$, whereas the IP-efficiency 
at maximum erasure rate is $1/2$.

The paper is organized as follows. In Sec. \ref{sec2}, we explain the notion of an information reservoir using a two-state version of the MJ model. A general inequality, from which the standard entropy production of stochastic 
thermodynamics and a novel IP-entropy production accounting for information reservoirs are obtained, is proved in Sec. \ref{sec3}.
Furthermore, with the transition rates fulfilling a generalized detailed balance relation, we identify the general first and second law for a thermodynamic system interacting with information reservoirs. 
In Sec. \ref{sec4}, we study a simple three-state model illustrating how an information
reservoir changes the second law and a two-state system that only interacts with information reservoirs, with no heat dissipation or work exchange. 
A linear response theory including information processing is developed in Sec. \ref{sec5}. We conclude in Sec. \ref{sec6}. 
In appendix \ref{appa}, we prove a master fluctuation theorem generalizing the inequality from Sec. \ref{sec3}.

\section{Paradigmatic model}
\label{sec2}

\subsection{Description of the model}

We can motivate our generalization of stochastic thermodynamics and give a clear interpretation of an information reservoir by starting with 
a simple paradigmatic model \cite{bara14}, which corresponds to a reduced (from six to two states) 
version of the MJ model. The system consists of two states, labeled $d$ and $u$. State $d$ has internal energy $0$ and the internal energy of state $u$ is $E$. Transitions between the states are mediated 
by thermal reservoir at temperature $k_BT=1$, implying 
\begin{equation}
k_+/k_-= \exp(-E),
\end{equation}
where $k_+$ is the transition rate from $d$ to $u$ and $k_-$ is the transition rate from $u$ to $d$. The system is also connected to a work reservoir. 

\begin{figure}
\includegraphics[width=72mm]{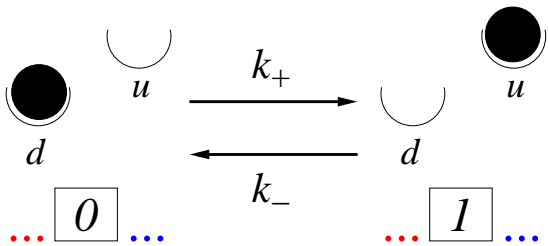}
\vspace{-2mm}
\caption{Two-state system interacting with a bit. If a thermal transition happens from $d$ ($u$) to $u$ ($d$) the bit changes its state
from $0$ ($1$) to $1$ ($0$). }
\label{fig1} 
\end{figure}

\begin{figure}
\includegraphics[width=82mm]{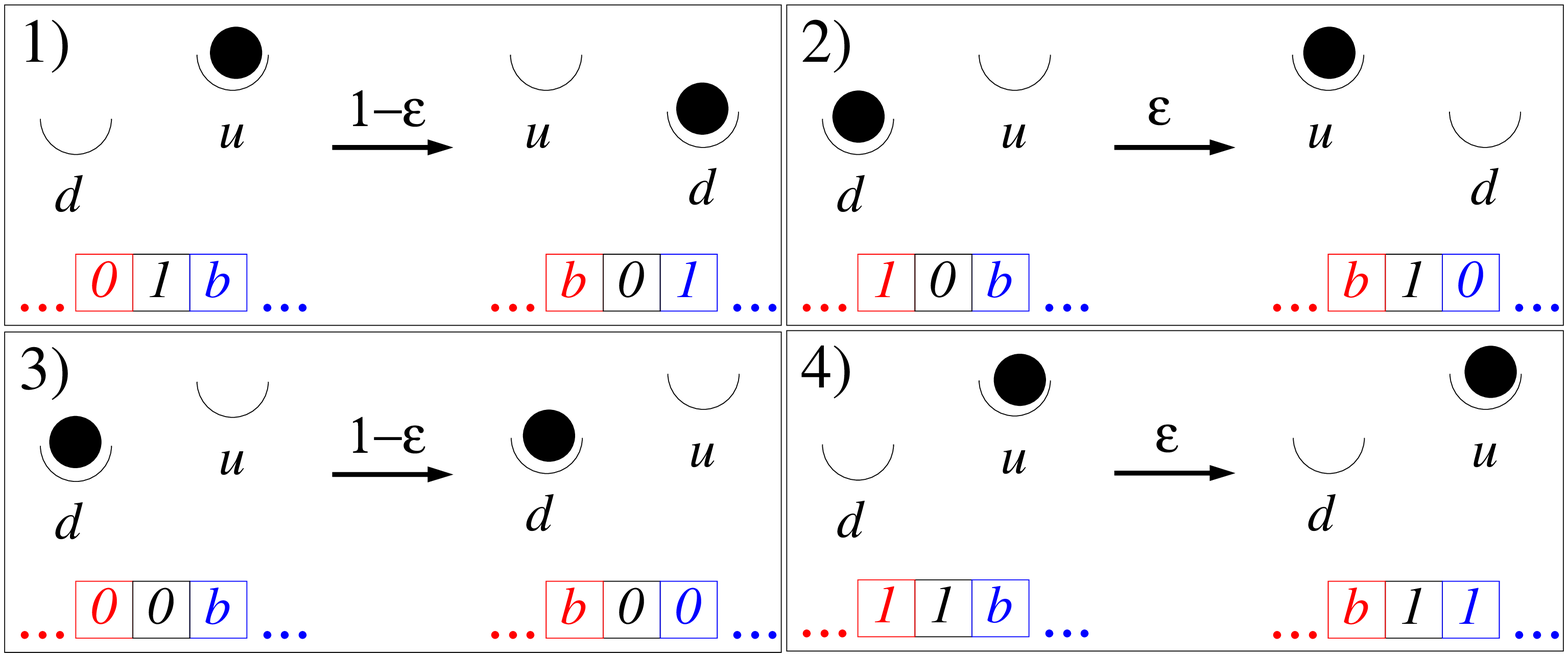}
\vspace{-2mm}
\caption{Possible effective transitions generated by the new incoming bit. Case 1 corresponds to the system being in state $u$ and the new incoming bit in state $0$, leading 
to extracting a quantity $E$ of work. In case 2 the system is in state $d$ and the new incoming bit is in state $1$, which leads to a quantity $E$ of work 
entering the system from the work reservoir. Case 3 (4) corresponds to the system being in state $d$ ($u$) and the the new incoming bit in state $0$ ($1$), which does
involve exchange of energy with the work reservoir. The letter $b$ in the tape represents a bit that can be either in state $0$ or $1$.   
 }
\label{fig2} 
\end{figure}

In order to extract work from a single heat bath an information reservoir is also needed, which can be understood as a sequence
of bits, i.e., a tape, that interacts with the system. As represented in Fig. \ref{fig1}, a bit from the tape interacts with the system for a certain time interval in such a way that the bit state $0$ ($1$) 
is coupled to the system state $d$ ($u$). For example, during this time interval, if the system makes a thermal 
transition from $d$ to $u$ the bit changes from $0$ to $1$. After this interaction time interval the tape moves one step forward, 
with the bit that interacted with the system leaving and a next bit from the tape coming 
to interact with the system. This new incoming bit can generate effective transitions between the states of the system, leading to an exchange of energy with the work reservoir, as shown in Fig. \ref{fig2}.

More precisely, if the system finishes the time interval 
in state $u$ and the new incoming bit is in state $0$, the energy levels of the system are interchanged with the occupied level $u$ being lowered to energy $0$ and the empty level $d$ being raised to energy $E$. 
The lowering of the occupied level $u$ leads to a work extraction of $E$. After changing the energy levels,  the labels of the states are also interchanged and, therefore, this operation leads to a transition from $u$ to $d$. 
In the same way, if the system finishes the time interval in state $d$ and the new incoming bit is in state $1$, then an effective transition from $d$ to $u$ resulting in work $E$ flowing from
the work reservoir to the system occurs. In the other two cases, namely, the system finishing the time interval in state $d$ and the new incoming bit being $0$ or the system finishing in state $u$ and the new incoming bit being $1$,
no work exchange takes place. 

\begin{figure}
\includegraphics[width=72mm]{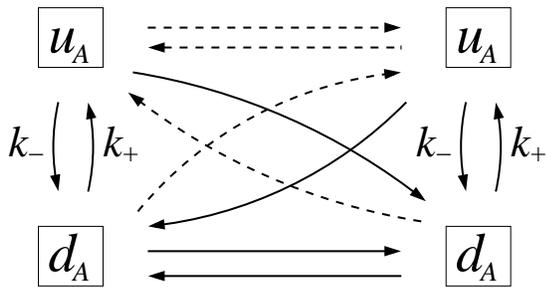}
\vspace{-2mm}
\caption{Transition rates for the four-state model. The thermal transitions take place with rates $k_+$ and $k_-$. Transition between states with a different subscript are related to the tape moving forward and a new 
bit coming to interact with the system. The solid arrows represent transitions with rate $\gamma(1-\epsilon$) and the dashed arrows with $\gamma\epsilon$.}
\label{fig3} 
\end{figure}

The probability that the new incoming bit is in state $1$ is $\epsilon$ and in state $0$ is $1-\epsilon$.    
The interaction time interval is assumed to be exponentially distributed with a rate $\gamma$, which characterizes the velocity of the tape. Assuming a constant time interval, as in \cite{mand12},
does not change the qualitative behavior of the model \cite{bara14}. The advantage of working with exponentially distributed time intervals is that the model can be described as a nonequilibrium steady state (NESS). 
The transition rates for the four-state Markov process, corresponding to a duplication of the two-state system are displayed in Fig. \ref{fig3}.
This duplication is necessary to include transitions generated by the new incoming bit. More precisely, 
a transition between the different subscripts $A$ and $B$ is generated by the new incoming bit and implies the tape moving forward. 
The time-scale for these transitions is then $1/\gamma$ and, as the new incoming bit does not depend on the state of the system, 
the transition rates between states with different subscripts are independent of the state of the system, e.g., the transition rate from $u_A$ to $d_B$ is the 
same as the transition rate from $d_A$ to $d_B$. Transitions between states with the same subscript are related to the thermal reservoir. 

\subsection{Work and Shannon entropy difference}

In the limit $k_+,k_-\gg\gamma$, the probability of finishing the interaction time interval in state $u$ is $p\equiv 1/(1+\exp E)$.
We denote the stationary probability of, for example, state $u_A$ as $P_{u_A}$.
Defining $\tau\equiv k/(k+\gamma)$, with $k\equiv k_++k_-$, the stationary probability of state $u$, in the four-state
model in Fig. \ref{fig3}, is 
\begin{equation}
p_\tau= \tau p+ (1-\tau)\epsilon,
\label{ptaueq}
\end{equation}
where $p_\tau\equiv P_{u_A}+P_{u_B}$. This stationary probability corresponds to the probability of finishing an interacting time interval in state $u$. In other words,
$p_\tau$ is the probability of being in state $u$ before a jump between different subscripts occurs.

The rate of extracted work is  
\begin{equation}
\dot{w}_{\textrm{out}}= \gamma E[p_\tau(1-\epsilon)-(1-p_\tau)\epsilon]=\gamma E[p_\tau-\epsilon].
\label{wout2s}
\end{equation}
Since $p_\tau$ is the probability of being in state $u$ at the end of an interacting time interval, the probability 
of finding a $1$ in the outgoing tape, which amounts to the sequence of bits
that has already interacted with the system, is $p_\tau$. This outgoing
tape is then a record of the interaction with the system, and has Shannon entropy 
\begin{equation}
H(p_\tau)\equiv -p_\tau\ln p_\tau-(1-p_\tau)\ln(1-p_\tau),
\end{equation}
while the incoming tape has Shannon entropy $H(\epsilon)$.
As we demonstrate in the next section, the following second law inequality holds,
\begin{equation}
\dot{s}_1= \gamma[H(p_\tau)-H(\epsilon)]-\dot{w}_{\textrm{out}}\ge0,
\label{s1a}
\end{equation}
where $\dot{s}_1$ is the IP-entropy production. The physical meaning of the inequality is the following. Let us consider the case $p\le 1/2$ and $\epsilon\le 1/2$.
For $\epsilon<p$, the system operates as a machine, with the extracted work being bounded by the Shannon entropy change 
in the tape $H(p_\tau)-H(\epsilon)$, which is positive. Considering a tape with larger Shannon entropy as containing more information, the capacity of the tape to store information is the thermodynamic resource that is consumed in this process.
If $\epsilon>p$, it is convenient to rewrite (\ref{s1a}) as
\begin{equation}
\dot{s}_1= \dot{w}-\gamma[H(\epsilon)-H(p_\tau)\ge0,
\label{s1b}
\end{equation}
where $\dot{w}= -\dot{w}_{\textrm{out}}$ is the rate of work entering the system. In this case the system operates as an eraser:
work is consumed in order to decrease the Shannon entropy of, or erase information from, the tape. For a complete discussion of 
the full phase diagram of a similar model see \cite{mand12}.
We note that the exactly same model can be interpreted as a feedback
driven system with a controller performing measurements.
With this interpretation a different entropy production is obtained \cite{bara14}.

\subsection{Reduction to a two-state model}

\begin{figure}
\includegraphics[width=72mm]{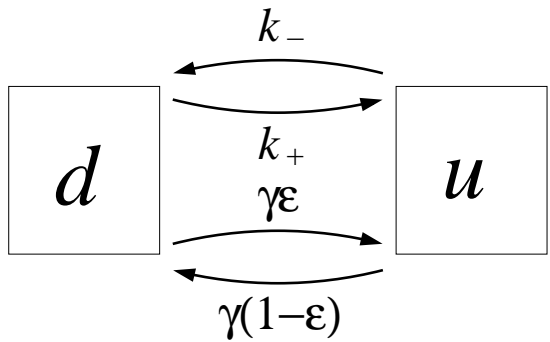}
\vspace{-2mm}
\caption{Two-state reduction of the four-state model from Fig. \ref{fig3}. }
\label{fig4} 
\end{figure}

The stationary state properties of the four-state model are identical to the stationary state properties of the two-state model represented in Fig. \ref{fig4}, with the stationary probability of state $u$ in the two-state model
$P_u=P_{u_A}+P_{u_B}$. 
Within this reduced two-state model, one link, with the transition rates $k_+$ and $k_-$, is related to a thermal reservoir. The other transitions are generated by
the information reservoir as explained above. Whenever the system makes a transition through the thermal link, heat is exchanged with the heat reservoir. If the transition
is through the link associated with the information reservoir the system exchanges work with the work reservoir.
From the first law the heat taken from the thermal reservoir equals the extracted work. The contribution to $\dot{s}_1$ in Eq. (\ref{s1a}) related to the link associated with the thermal reservoir 
is the rate of dissipated heat $-\dot{w}_{\textrm{out}}$ and the contribution of the link associated with the information reservoir is $\gamma(H(p_\tau)-H(\epsilon))$.

As we will show in the next sections a more general second law inequality allows for this interpretation of any link between states as being associated with an information reservoir.
The terms in $\dot{s}_1$ related to information reservoirs are proportional to a Shannon entropy change, as is $H(p_\tau)-H(\epsilon)$ in (\ref{s1a}). 

\subsection{Relation with the standard entropy production}

Besides $\dot{s}_1$, the standard thermodynamic entropy production of stochastic thermodynamics \cite{seif12} for the two-state model reads
\begin{equation}
\dot{s}= \gamma(p_\tau-\epsilon)\ln\frac{1-\epsilon}{\epsilon}-\dot{w}_{\textrm{out}}\ge0.
\label{sa}
\end{equation}
Comparing with the entropy rate (\ref{s1a}) we obtain $\dot{s}\ge\dot{s_1}$. 
The contribution  $\gamma(p_\tau-\epsilon)\ln\frac{1-\epsilon}{\epsilon}$ has a clear physical interpretation. Let us first take $p<\epsilon$. 
Consider another two-state system with which we can reset the tape. The energy difference of this auxiliary system is chosen as $E'=\ln[(1-\epsilon)/\epsilon]$, 
the incoming tape is characterized by the probability of a $1$ being $p_\tau$, and $k'\gg \gamma$, where $k'$ is the time-scale of its
thermal transitions. The auxiliary two-state system acts as an eraser and its entropy rate (\ref{s1b}) becomes
\begin{equation}
\dot{s}_1'= \gamma(p_\tau-\epsilon)\ln\frac{1-\epsilon}{\epsilon}-\gamma[H(p_\tau)-H(\epsilon)]\ge0,
\end{equation}
where the first term  is obtained from (\ref{wout2s}) with energy $E'=\ln[(1-\epsilon)/\epsilon]$. 
Hence, the term $\gamma(p_\tau-\epsilon)\ln\frac{1-\epsilon}{\epsilon}$, appearing in (\ref{sa}) is the rate of work that must be consumed, which equals the rate of heat that must be dissipated, in order to recover the original tape 
with Shannon entropy $H(\epsilon)$ from a tape with Shannon entropy $H(p_\tau)$, using an auxiliary two-state system with $k'\gg \gamma$ and $E'=\ln[(1-\epsilon)/\epsilon]$.

Similarly, if $p<\epsilon$, the term $\gamma(\epsilon-p_\tau)\ln\frac{1-\epsilon}{\epsilon}$ in $\dot{s}$ corresponds to the work that would be extracted from an incoming tape with Shannon entropy $H(p_\tau)$ interacting with
the auxiliary two-state system with $E'=\ln[(1-\epsilon)/\epsilon]$ and  $k'\gg \gamma$. Hence, the standard entropy production of stochastic thermodynamic $\dot{s}$ contains the full thermodynamic cost of restoring the tape
to its original distribution \cite{bara14}.

\section{General theory}

\label{sec3}

\subsection{First and second law}

We consider a thermodynamic system with generic states denoted by $i$ and $j$ with internal energy $E_i$ and $E_j$. This system is in contact with reservoirs $\nu$ at inverse temperature $\beta_\nu$.
In a transition from $i$ to $j$, besides exchanging heat with the reservoir $\nu$ the system can also exchange work if a generic quantity $d_{ij}^\alpha$ changes. The field associated with this quantity and 
reservoir $ \nu$ is $f^\alpha_\nu$. For example, if $d^{\alpha}_{ij}= N_j-N_i$, where $N_i$ is the number of particles in the system in state $i$, then $f^\alpha_\nu= \mu_\nu$ is the chemical potential 
of these particles. Note that changing the parameter $\nu$ corresponds to a different chemical potential and the same particles, whereas, changing $\alpha$ could correspond to another kind of particle that is
exchanged with the system.

\begin{figure}
\includegraphics[width=72mm]{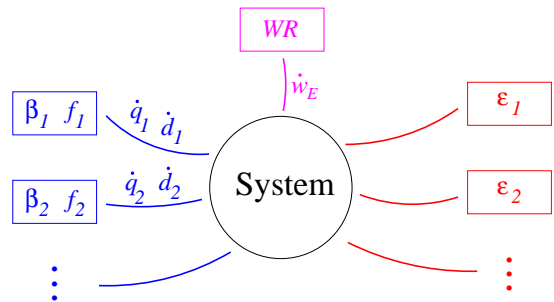}
\vspace{-2mm}
\caption{Sketch of a system interacting with standard reservoirs with inverse temperature $\beta_\nu$ and field $f_\nu$. Information reservoirs are characterized  by $\epsilon_n$, the probability of a bit in state $1$.
The additional work reservoir, related to transitions mediated by the information reservoir for which the internal energy of the system changes,  is indicated by $WR$.}
\label{fig6} 
\end{figure}

Besides these standard reservoirs the system also interacts with information reservoirs, which can be understood as a tape interacting with a pair of states of the system in the way explained in Sec. \ref{sec2}. An information
reservoir $n$ is characterized by $\epsilon_n$, the probability that an incoming bit is in the state $1$.
The coupling between information reservoirs and the system changes the entropy balance while keeping the first law intact, as they do not exchange energy with the system. In Fig. \ref{fig6}, a system interacting with both
standard and information reservoirs is depicted. There is also an additional work reservoir, which is related to the fact that  if the system goes 
from state $i$ to $j$ through a transition mediated by an information reservoir the change in internal energy of the system is assumed to be compensated by an exchange of work with 
this additional work reservoir.      
  
The system is assumed to be described by Markovian dynamics with the transition rates from $i$ to $j$ related to a standard reservoir $\nu$ being $W_{ij}^{(\nu)}$.  
These transition rates fulfill the local detailed balance relation \cite{seif11a}
\begin{equation}
\ln \frac{W_{ij}^{(\nu)}}{W_{ji}^{(\nu)}}= -\beta_\nu(E_j-E_i)+\beta_\nu\sum_\alpha f^\alpha_\nu d_{ij}^\alpha.
\end{equation}
For an information reservoir $n$, the associated transition rates fulfill
\begin{equation}
\ln \frac{W_{ij}^{(n)}}{W_{ji}^{(n)}}= \ln\frac{\epsilon_n}{1-\epsilon_n},
\label{infaff}
\end{equation}
where state $i$ is related to the bit state $0$ and state $j$ to $1$.

The steady state probability of state $i$ is denoted $P_i$ and the stationary probability current from $i$ to $j$ related to reservoir $\xi$ is 
\begin{equation}
J_{ij}^{(\xi)}\equiv P_iW_{ij}^{(\xi)}-P_jW_{ji}^{(\xi)},
\label{gencurr}
\end{equation} 
where $\xi$ can be either a standard or an information reservoir. The rate of internal energy variation related to transitions mediated by reservoir $\xi$ is  
\begin{equation}
\dot{E}_\xi\equiv \sum_{i<j}J_{ij}^{(\xi)}(E_j-E_i),
\end{equation}
where $\sum_{i<j}$ means a sum over all pairs $ij$ without summing the same pair twice. 
In the steady state, the contribution due to all reservoirs must be zero, i.e.,
\begin{equation}
\dot{E}\equiv \sum_\xi\dot{E}_\xi= 0.
\end{equation}
Furthermore, the rate of variation of a generic quantity $d_{ij}^{\alpha}$ due to the interaction with a standard reservoir $\nu$ reads 
\begin{equation}
\dot{d}_\nu^\alpha= \sum_{i<j}J_{ij}^{(\nu)}d_{ij}^\alpha.
\end{equation}
The rate of heat dissipated in reservoir $\nu$ is identified as
\begin{equation}
\dot{q}_\nu= -\dot{E}_\nu+ \sum_{\alpha}f^\alpha_\nu\dot{d}_\nu^\alpha.
\label{heatdef}
\end{equation}
Information reservoirs $n$, on the other hand, do not involve any heat dissipation. 

The rate of work entering the system is 
\begin{equation}
\dot{w}\equiv \sum_{\nu,\alpha} f^\alpha_\nu\dot{d}^\alpha_\nu+\sum_n\dot{E}_n\equiv \sum_{\nu,\alpha} f^\alpha_\nu\dot{d}^\alpha_\nu+\dot{w}_E,
\end{equation}
where the contribution $\dot{w}_E= \sum_n\dot{E}_n$ is the work entering the system from the additional work reservoir.
The first law then becomes
\begin{equation}
\dot{E}= -\sum_\nu\dot{q}_\nu+\dot{w}=0.
\label{firstlaw}
\end{equation}

The second law inequality generalizing stochastic thermodynamics for a system interacting with information reservoirs, which follows from a more general inequality proved 
in the next subsection, reads
\begin{equation}
\dot{s}_1= \sum_\nu\beta_\nu \dot{q}_\nu+\sum_n\dot{h}_n\ge 0
\label{entinfphys}
\end{equation}
where 
\begin{equation}
\dot{h}_n\equiv\sum_{i<j}\gamma_{ij}^{(n)}[H(p_{ij})-H(\epsilon_n)],
\label{sharate}
\end{equation}
with 
\begin{equation}
\gamma_{ij}^{(n)}\equiv(P_i+P_j)(W_{ij}^{(n)}+W_{ji}^{(n)})
\label{gammadef}
\end{equation}
and
\begin{equation}
p_{ij}\equiv P_j/(P_i+P_j).   
\label{pijdef}
\end{equation}
The term $\dot{h}_n$ is the rate at which the entropy of the information reservoir changes due to the interaction with the system. The term  $\gamma_{ij}^{(n)}$ in Eq. (\ref{sharate})
is the time-scale for transitions between $i$ and $j$ through $n$ multiplied by the stationary probability of the pair of states $P_i+P_j$, whereas the term $H(p_{ij})-H(\epsilon_n)$ is the Shannon entropy
change, with the outgoing tape being a record of the stationary relative probability of the pair $p_{ij}$.  

If an information reservoir labeled by $n$ is related to more than one pair of states, one can imagine that each pair is related to a different tape, with all incoming tapes having distribution $\epsilon_n$ and each outgoing 
tape having distribution $p_{ij}$. The information reservoir does not need to be understood as a tape of ordered bits running through the system. Another possibility is to consider it 
as some bath of particles that can be in states $0$ and $1$ \cite{bara13}. During a transition, the system takes a new particle from
this bath with distribution $\epsilon_n$ and releases the old particle to another bath that will have distribution $p_{ij}$. Within this view, the same 
bath is related to all pair of states associated with $n$.

We note that it is also possible to study entropic interactions with the standard entropy production.  
Specifically, assigning an intrinsic entropy to a state $i$ \cite{seif11,espo12}, entropic currents related to this intrinsic entropy appear in the standard 
thermodynamic entropy production, modifying the second law while keeping the first law unaltered. Entropic currents can also be interpreted as being related 
to a Maxwell's demon monitoring the transitions of the system \cite{espo12b}. Moreover, in a recent case study of a quantum dot interacting with a tape,
the term related to the Shannon entropy change was found to be proportional to an entropic current \cite{stra14}.

\subsection{Proof of the generalized second law-like inequality}

Assuming first that there is only one link for each pair of states, the stationary master equation reads
\begin{equation}
\sum_{j\neq i}[P_jW_{ji}-P_iW_{ij}]=0.
\label{meq}
\end{equation}
The standard thermodynamic entropy production is 
\begin{equation}
\dot{s}\equiv \sum_i\sum_{j\neq i}P_i W_{ij}\ln\frac{W_{ij}}{W_{ji}}\ge0.
\label{secine}
\end{equation}
To obtain a more general formula, we consider auxiliary transition rates $\overline{W}_{ij}$.
Moreover, we define the quantities $R_{i}\ge0$ and $\overline{R}_{i}\ge 0$, 
which are constrained to fulfill the relation
\begin{equation}
R_{i}+\sum_{j\neq i}W_{ij}=\overline{R}_{i}+\sum_{j\neq i}\overline{W}_{ij}.
\label{Wconstraint}
\end{equation}
With these auxiliary transition rates we define 
\begin{equation}
\dot\omega\equiv \sum_i\left(\sum_{j\neq i}P_iW_{ij}\ln\frac{W_{ij}}{\overline{W}_{ji}}+P_iR_{i}\ln\frac{R_{i}}{\overline{R}_{i}}\right)
\label{eqom}
\end{equation}
Using the inequality $-\ln x\ge1-x$ and summing $\sum_i\sum_{j\neq i}P_i W_{ij}\ln(P_i/P_j)=0$ to the right hand side of the above equation,
we obtain
\begin{eqnarray}
\dot\omega\ge \sum_i\sum_{j\neq i}(P_iW_{ij}-P_j\overline{W}_{ji})+\sum_iP_i(R_{i}-\overline{R}_{i})\nonumber\\
=\sum_i\sum_{j\neq i}(P_i\overline{W}_{ij}-P_j\overline{W}_{ji})=0,
\label{gensecond}
\end{eqnarray}
where we used Eq. (\ref{Wconstraint}). This inequality is a generalization of (\ref{secine}), since for the choice $\overline{W}_{ij}= W_{ij}$ the
rate $\dot{\omega}$ becomes the entropy production $\dot{s}$. A fluctuation theorem generalizing (\ref{gensecond}) is proved in App. \ref{appa}.

We now consider the possibility of more than one link between the same pair of states, since different reservoirs can be related to the same pair of states. This is the case of the two-state model of Sec. \ref{sec2}.  
In this case the total transition rate reads $W_{ij}= \sum_\xi W_{ij}^{(\xi)}$, where $\xi$ label different links (reservoirs). The same is valid for the auxiliary rates $\overline{W}_{ij}= \sum_\xi \overline{W}_{ij}^{(\xi)}$. Furthermore, 
for convenience, we write $R_i= \sum_{j\neq i}\sum_\xi R_{ij}^{(\xi)}$ and $\overline{R}_i= \sum_{j\neq i}\sum_\xi \overline{R}_{ij}^{(\xi)}$. For multiple reservoirs  we then define the quantity  
\begin{equation}
\dot\omega'\equiv \sum_i\sum_{j\neq i}\sum_\xi\left( P_iW^{(\xi)}_{ij}\ln\frac{W^{(\xi)}_{ij}}{\overline{W}^{(\xi)}_{ji}}+P_iR_{ij}^{(\xi)}\ln\frac{R_{ij}^{(\xi)}}{\overline{R}_{ij}^{(\xi)}}\right)\ge 0,
\label{gensecond2}
\end{equation}
which, from the log sum inequality, is larger than $\dot{\omega}$ defined in Eq. (\ref{eqom}).

The standard entropy production with multiple links becomes \cite{seif12}
\begin{equation}
\dot{s}\equiv \sum_{i<j}\sum_\xi J_{ij}^{(\xi)}\mathcal{F}_{ij}^{(\xi)},
\label{fullent}
\end{equation}
where $\mathcal{F}_{ij}^{(\xi)}\equiv \ln(W_{ij}^{(\xi)}/W_{ji}^{(\xi)})$. This formula can also be obtained from Eq. (\ref{gensecond2}) by setting $W_{ij}^{(\xi)}=\overline{W}_{ij}^{(\xi)}$ and $R_i=\overline{R_i}$. 
To obtain the IP-entropy production we separate the links $\xi$ into links related to standard reservoirs $\nu$ and link related to information reservoirs $n$. For the $\nu$ links the choice for the auxiliary 
rates is the same as the one used to obtain $\dot{s}$. For reservoirs $n$, choosing $\overline{W}_{ij}^{(n)}= p_{ij}(W_{ij}^{(n)}+W_{ji}^{(n)})$, $\overline{W}_{ji}^{(n)}= (1-p_{ij})(W_{ij}^{(n)}+W_{ji}^{(n)})$, 
$R_{ij}^{(n)}=W_{ji}^{(n)}$, and $\overline{R}_{ij}^{(n)}=\overline{W}_{ji}^{(n)}$, Eq. (\ref{gensecond2}) becomes the IP-entropy production
\begin{equation}
\dot{s}_1= \sum_{i<j}\left(\sum_\nu J_{ij}^{(\nu)}\mathcal{F}_{ij}^{(\nu)}\right)+\sum_n \dot{h}_n,
\label{infent} 
\end{equation}
where $\dot{h}_n$ is defined in Eq. (\ref{sharate})

From (\ref{fullent}) and (\ref{infent}), we obtain the difference between $\dot{s}$ and $\dot{s}_1$ as  
\begin{equation}
\dot{s}-\dot{s}_1= \sum_{i<j}\sum_n \gamma_{ij}^{(n)}D_{KL}(p_{ij}||\epsilon_n)\ge0,
\end{equation}
where 
\begin{equation}
D_{KL}(p_{ij}||\epsilon_n)\equiv p_{ij}\ln\frac{p_{ij}}{\epsilon_n}+(1-p_{ij})\ln\frac{(1-p_{ij})}{(1-\epsilon_n)}\ge 0
\end{equation}
is the Kullback-Leibler distance \cite{cove06}.
The physical meaning of this inequality is the same as in the two-state model.  The standard 
entropy production $\dot{s}$ contains the thermodynamic cost of resetting each tape $n$, using an auxiliary two-state system as discussed in Sec. \ref{sec2}.

\section{Further Examples}
\label{sec4}

\subsection{Refrigerator powered by a tape}
\label{sec4a}

In the model analyzed in Sec. \ref{sec2}, the presence of an information reservoir allowed the work extraction from a single heat bath. Using inequality (\ref{entinfphys}), we now introduce
a simple model where the presence of a information reservoir allows heat to flow from a cold to a hot reservoir. A four-state model with fixed interaction time intervals for
a refrigerator powered by a tape has been analyzed in \cite{mand13}.   

\begin{figure}
\includegraphics[width=62mm]{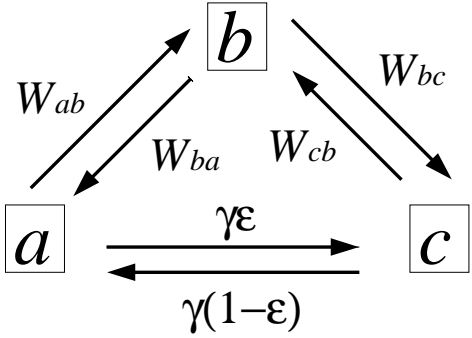}
\vspace{-2mm}
\caption{Three-state model. The rates $W_{ac}=\gamma\epsilon$ and are $W_{ca}=\gamma(1-\epsilon)$ are relate to the information reservoir. They are
the same for both the refrigerator powered by a tape (Sec. \ref{sec4a}) and the thermoelectric machine interacting with a tape (Sec. \ref{sec4b}).}
\label{fig5} 
\end{figure}

For a system interacting with  one information reservoir, related to a rate of Shannon entropy change $\dot{h}$ from Eq. (\ref{sharate}), and two heat reservoirs at inverse temperatures $\beta_1$ and $\beta_2$, with $\beta_2\le\beta_1$,
the first law (\ref{firstlaw}) becomes 
\begin{equation}
\dot{q}_2=-\dot{q}_1\equiv\dot{q},
\end{equation} 
where $\dot{q}$ is the rate at which heat flows from the cold to the hot reservoir defined in (\ref{heatdef}). The IP-entropy production (\ref{entinfphys}) is
\begin{equation}
\dot{s}_1=\dot{h}-\dot{q}(\beta_1-\beta_2)\ge0.
\end{equation} 
Hence, if $\dot{h}\ge 0$ then $\dot{q}$ can be positive, i.e., heat can flow from the cold to the hot reservoir. This specific form of the second law 
has also been obtained in \cite{deff13} using an Hamiltonian formalism.

A specific  three-state model with states $a$, $b$, and $c$ is represented in Fig. \ref{fig5}. The transition rates between $a$ and $b$ are associated with the cold reservoir at inverse temperature $\beta_1$, whereas the 
transition rates between $b$ and $c$ are associated with the hot reservoir with inverse temperature $\beta_2\le\beta_1$. States $a$ and $c$ have internal energy $0$, and state $b$ has internal energy $E$. The
local detailed balance relation then reads
\begin{equation}
\ln\frac{W_{ab}}{W_{ba}}= -\beta_1E\qquad\textrm{and}\qquad\ln\frac{W_{bc}}{W_{cb}}= \beta_2E.
\end{equation}   
We choose these transition rates as $W_{ab}=k \textrm{e}^{-E\beta_1/2}$, $W_{ba}=k \textrm{e}^{E\beta_1/2}$, $W_{bc}=k \textrm{e}^{E\beta_2/2}$, and $W_{cb}=k \textrm{e}^{-E\beta_2/2}$. The parameter $k$ sets the
time-scale of the thermal transitions. 

The transition rates between $a$ and $c$ are related to an information reservoir such that state $a$ ($c$) is coupled to the bit state $0$ ($1$). With the probability of a bit in state $1$ being $\epsilon\le 1/2$ in the incoming
tape, the transition rates are then written as $W_{ac}= \gamma \epsilon$ and $W_{ca}= \gamma(1-\epsilon)$, where $\gamma$ sets the time-scale of the information reservoir.


Calculating the stationary probability distribution we obtain 
\begin{equation}
p_\tau\equiv \frac{P_c}{P_a+P_c}=\frac{C_1p\tau+C_2\epsilon(1-\tau)}{C_1\tau+C_2(1-\tau)},
\end{equation}
where $C_1\equiv \textrm{e}^{\beta_2E}+\textrm{e}^{\beta_1E}$, $C_2\equiv \textrm{e}^{(\beta_1+\beta_2)E/2}(\textrm{e}^{\beta_2E/2}+\textrm{e}^{\beta_1E/2})$, and $\tau\equiv k/(k+\gamma)$. Furthermore, the probability current in the clockwise direction in Fig. \ref{fig5} is
\begin{equation}
J\equiv \gamma[(1-\epsilon)P_c-\epsilon P_a]= \gamma(P_a+P_c)[p_\tau-\epsilon]\propto (p-\epsilon),
\label{curr3}
\end{equation}
where 
\begin{equation}
p\equiv \lim_{\tau\to 1} p_\tau= \frac{1}{1+\textrm{e}^{(\beta_1-\beta_2)E}}\le1/2.
\end{equation}
Restricting to $\epsilon\le 1/2$, for $p>\epsilon$ the probability current in Eq. (\ref{curr3}) is positive leading to heat flowing from the cold to the hot reservoir.
More precisely, the IP-entropy production (\ref{entinfphys}) becomes
\begin{equation}
\dot{s}_1= \gamma(P_a+P_c)[H(p_\tau)-H(\epsilon)]-\dot{q}(\beta_1-\beta_2),
\end{equation}
where $\dot{q}= JE$ is the rate at which heat flows from the cold to the hot reservoir. The refrigerator mode of operation ($p>\epsilon$) 
is powered by the tape, which has its Shannon entropy increased from $H(\epsilon)$ to $H(p_\tau)$. For $p<\epsilon$ the probability current $J$ becomes negative and heat flows from 
the hot to the cold reservoir. In this case information is erased from the tape and the rate of Shannon entropy decrease of the tape is compensated by the rate of entropy increase of the external environment due to the
heat flow, i.e., $\gamma(P_a+P_c)[H(\epsilon)-H(p_\tau)]\le-\dot{q}(\beta_1-\beta_2)$.

\subsection{Thermoelectric machine interacting with a tape}
\label{sec4b}

We now consider the case where the system also exchanges particles with the standard reservoirs. The system is in contact with a reservoir at inverse temperature $\beta_1$ and chemical potential $\mu_1$,
another reservoir characterized by $\beta_2$ and $\mu_2$, and an information reservoir. The chemical potentials fulfill $\Delta\mu\equiv \mu_2-\mu_1\ge0$, where  $1$ is assumed to be the hot reservoir, i.e.,
$\beta_2\ge\beta_1$. The first law (\ref{firstlaw}) is reduced to 
\begin{equation}
-\dot{q}_2-\dot{q}_1=-\dot{w}.
\end{equation} 
where $-\dot{w}=\dot{N}\Delta \mu$ is the rate of work extracted from the system to move particles against the chemical potential gradient $\Delta \mu$ (from $1$ to $2$) at a rate $\dot{N}$,
and $\dot{q}_1$ ($\dot{q}_2$) is the rate of dissipated heat related to reservoir $1$ ($2$).  
The IP-entropy production (\ref{entinfphys}) for this case reads 
\begin{equation}
\dot{s}_1=\dot{h}+\beta_1\dot{q}_1+\beta_2\dot{q}_2\ge0,
\end{equation}
where $\dot{h}$ is the rate of Shannon entropy change given in Eq. (\ref{sharate}).
First, we note that if $\beta_1=\beta_2$ a positive $\dot{h}$ can move particles against the chemical potential. This corresponds to extracting work from a single heat bath, which was also the case
of the model from Sec. \ref{sec2}. Second, for the case where the temperature gradient $\beta_2-\beta_1$ drives the particles against $\Delta \mu$,
the pseudo-efficiency 
\begin{equation}
\eta_{\textrm{ps}}\equiv -\dot{w}/(-\dot{q}_1)
\label{pseudoeff}
\end{equation}
becomes
\begin{equation}
\eta_{\textrm{ps}}\le \eta_c+\frac{\dot{h}}{\beta_2(-\dot{q}_1)}, 
\label{eefstrange}
\end{equation}
where $\eta_c\equiv 1-\beta_1/\beta_2$ is the Carnot efficiency. Hence this pseudo-efficiency can exceeded the Carnot efficiency $\eta_c$. Actually, it can even exceed $1$ as demonstrated below. A relation 
similar to (\ref{eefstrange}) has also been obtained in \cite{espo12b,deff13} using different frameworks.

As a specific model describing such situation we take the three-state model from Fig. \ref{fig5}. 
We now assume that in state $b$ the number of particles in the system is $N=1$ and in states
$a$ and $c$  it is $N=0$. The local detailed balance relation must be modified to
\begin{equation}
\ln\frac{W_{ab}}{W_{ba}}= -\beta_1(E-\mu_1)\qquad\textrm{and}\qquad\ln\frac{W_{bc}}{W_{cb}}= \beta_2(E-\mu_2),
\end{equation} 
where $\mu_1$ and $\mu_2$ are chemical potentials. We set these transition rates to 
$W_{ab}=k \textrm{e}^{-(E-\mu_1)\beta_1/2}$, $W_{ba}=k \textrm{e}^{(E-\mu_1)\beta_1/2}$, $W_{bc}=k \textrm{e}^{(E-\mu_2)\beta_2/2}$, and 
$W_{cb}=k \textrm{e}^{-(E-\mu_2)\beta_2/2}$. The transition rates between $a$ and $c$ are mediated by an information reservoir and are as in the model from Sec. \ref{sec4a}.

Calculating the stationary distribution we obtain
\begin{equation}
p_\tau\equiv\frac{P_c}{P_a+P_c}= \frac{C_1\tau p+C_2(1-\tau)\epsilon}{C_1\tau+C_2(1-\tau)},
\end{equation}    
where $\tau\equiv k/(k+\gamma)$, $C_1\equiv \textrm{e}^{\beta_2E+\beta_1\mu_1}+\textrm{e}^{\beta_1E+\beta_2\mu_2}$, 
$C_2\equiv \textrm{e}^{(\beta_1+\beta_2)E/2}(\textrm{e}^{\beta_2E/2+\beta_1\mu_1/2}+\textrm{e}^{\beta_1E/2+\beta_2\mu_2/2})$, and
\begin{equation}
p\equiv 1/(1+\textrm{e}^{\beta_2[(\mu_2-\mu_1)+\eta_c(\mu_1-E)]}).
\label{defp}
\end{equation}
The probability current is again
\begin{equation}
J\equiv \gamma[(1-\epsilon)P_c-\epsilon P_a]= \gamma(P_a+P_c)[p_\tau-\epsilon]\propto (p-\epsilon).
\end{equation}
Therefore, the rate of heat taken from the hot reservoir becomes
\begin{equation}
-\dot{q}_1= (E-\mu_1)J,
\label{eqq1}
\end{equation}
the rate of heat dissipated in the cold reservoir  
\begin{equation}
\dot{q}_2=(E-\mu_2)J,
\label{q2model}
\end{equation}
and the rate of extracted work
\begin{equation}
-\dot{w}=(\mu_2-\mu_1)J.
\label{eqw}
\end{equation}
Moreover, the rate at which the Shannon entropy of the information reservoir increases due to the interaction with the system is
\begin{equation}
\dot{h}=\gamma (P_a+P_c)[H(p_\tau)-H(\epsilon)].
\end{equation}

\begin{figure}
\includegraphics[width=72mm]{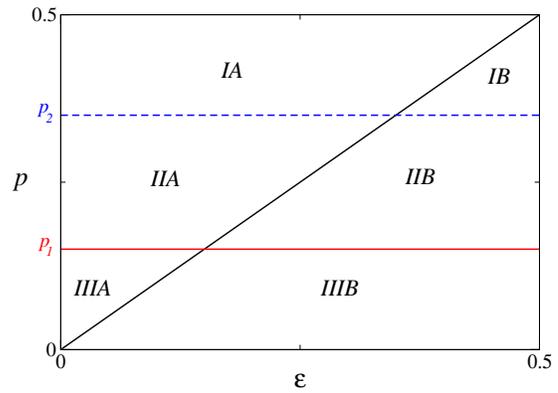}
\vspace{-2mm}
\caption{Phase diagram of the three-state model from Fig. \ref{fig5} with particle exchange with the reservoirs. The signs of the triplet $(\dot{q}_1,\dot{q}_2,\dot{w})$ are: $(-,+,-)$ in $IA$;
$(-,-,-)$ in $IIA$; $(+,-,-)$ in $IIIA$; $(+,-,+)$ in $IB$;
$(+,+,+)$ in $IIB$; $(-,+,+)$ in $IIIB$. The differences between the phases are explained in the text.
}
\label{fig7} 
\end{figure}

We restrict to the case $\epsilon\le 1/2$ and $p\le 1/2$, which from (\ref{defp}) implies $E\le (\beta_2\mu_2-\beta_1\mu_1)/(\beta_2-\beta_1)$. From Eq. (\ref{eqq1}) and Eq. (\ref{eqw}), the pseudo-efficiency (\ref{pseudoeff}) is
given by 
\begin{equation}
\eta_{\textrm{ps}}= (\mu_2-\mu_1)/(E-\mu_1).
\end{equation}
We define $p_2$ ($p_1$) as the probability $p$, given in Eq. (\ref{defp}), for $E=\mu_2$ ($E=\mu_1$).
The phase diagram of the model is shown in Fig. \ref{fig7}. First we take $p>\epsilon$, for which $\dot{h}\ge 0$. For $p>p_2$, corresponding to region $IA$ in Fig. \ref{fig7},
the pseudo-efficiency $\eta_{\textrm{ps}}$ is smaller than one
and the system operates as a standard thermoelectric  machine with an improved efficiency. In region $IIA$ with  $p<p_2$, the system takes heat from the hot and the cold reservoir, i.e., $\dot{q}_2$ in Eq. (\ref{q2model}) becomes negative. The
pseudo-efficiency then fulfills $\eta_{\textrm{ps}}> 1$, since the extracted work is larger than the heat taken from the hot reservoir. For $p\to p_1$ from above $\eta_{\textrm{ps}}\to \infty$. Crossing to region $IIIA$, where
$p<p_1$, the pseudo-efficiency becomes formally negative: the system takes heat from the cold reservoir, dissipates heat in the hot reservoir and does work against the chemical gradient. The unusual 
modes of operation $IIA$ and $IIIA$ are only possible because of the entropy increase in the information reservoir.

Second we consider $p<\epsilon$, corresponding to erasure of information from the tape. In the region $IB$ the system operates as a refrigerator, with the work entering the system $\dot{w}$ being used to erase the tape
and produce a heat flow from the cold to the hot reservoir. In the region $IIB$ the work entering the system is dissipated as heat in both reservoirs. In the region $IIIB$  
the system takes heat from the hot reservoir and dissipates heat in the cold reservoir.

\subsection{System interacting with two tapes}

It is also possible for a system to interact with more than one information reservoir. The simplest case is a system interacting with two information reservoirs, 
with no exchange of energy. As an example, we consider a two-state model with two links between the states as the model from Sec. \ref{sec2}. However, instead of one link 
being related to a thermal reservoir, both links are associated with information reservoirs. For one tape the probability of a $1$ is $\epsilon_1\le 1/2$ and for the other one 
this probability is $\epsilon_2\le 1/2$. The bit state $0$ ($1$) couples with state $d$ ($u$). The transition rates from $d$ to $u$ is $\gamma_1\epsilon_1$ for link $1$ and $\gamma_2\epsilon_2$ for link $2$.
The reversed transition rates from $u$ to $d$ are $\gamma_1(1-\epsilon_1)$ and $\gamma_2(1-\epsilon_2)$, respectively.  

The IP-entropy production (\ref{entinfphys}) is 
\begin{equation}
\gamma_1[H(p_\tau)-H(\epsilon_1)]+\gamma_2[H(p_\tau)-H(\epsilon_2)]\ge 0,
\end{equation}   
where $p_\tau\equiv (\gamma_1\epsilon_1+\gamma_2\epsilon_2)/(\gamma_1+\gamma_2)$. Assuming $\epsilon_1<\epsilon_2\le1/2$, information is written on tape $1$ and erased from tape $2$.
The efficiency of erasing information is   
\begin{equation}
\eta\equiv \frac{\gamma_2[H(\epsilon_2)-H(p_\tau)]}{\gamma_1[H(p_\tau)-H(\epsilon_1)]}\le 1.
\end{equation}
We call any efficiency involving a rate of Shannon entropy change of an information reservoir, as the efficiency above, an IP-efficiency. 
For $\gamma_2\gg \gamma_1$ we obtain 
\begin{equation}
\eta\to \frac{H(\epsilon_2)-H(\epsilon_1)-D_{KL}(\epsilon_1||\epsilon_2)}{H(\epsilon_2)-H(\epsilon_1)},
\end{equation}
and for $\gamma_2\ll \gamma_1$ the IP-efficiency reaches 
\begin{equation}
\eta\to \frac{H(\epsilon_2)-H(\epsilon_1)}{H(\epsilon_2)-H(\epsilon_1)+D_{KL}(\epsilon_2||\epsilon_1)}.
\end{equation}
It is interesting to compare the present situation with the case of a model in contact with two heat baths, for which heat flows from the hot to the cold reservoir.
For the system in contact with thermal reservoirs, the heat that leaves the hot reservoir is the heat entering the cold reservoir. On the other hand, information (or entropy), unlike energy,  is in general
not conserved, with the information erased from tape $2$ being smaller than the information written on tape $1$.

\section{Linear response theory}

\label{sec5}
\subsection{IP-entropy production within linear response}

We denote ordinary affinities by $\mathcal{F}_k$ and the conjugate flux by $J_k$. The number of independent ordinary affinities (or fluxes) depend on how many standard reservoirs $\nu$ and fields $f^\alpha_\nu$ 
we have. For example, for two reservoirs $\nu=1,2$ exchanging energy and particles, related to the chemical potentials $\mu_1$ and $\mu_2$, there are two ordinary affinities $k=I,II$. The first affinity is $\mathcal{F}_I= \beta_2-\beta_1$ and the 
associated flux is $J_I=\sum_{i<j}J_{ij}^{(1)}(E_j-E_i)$. The second affinity is $\mathcal{F}_{II}= \mu_2\beta_2-\mu_1\beta_1$ and the associated flux is $J_{II}=\sum_{i<j}J_{ij}^{(2)}(N_j-N_i)$.

For simplicity we assume that each information reservoir $n$ is related to only one pair $ij$ so that $\gamma_{ij}^{(n)}= \gamma_n$ and $p_{ij}= p_n$, 
where $\gamma_{ij}^{(n)}$ is defined in (\ref{gammadef}) and $p_{ij}$ in (\ref{pijdef}). The standard entropy production $\dot{s}$ is known to be given by a sum of terms composed by a current multiplying an affinity \cite{seif12}. Hence,
from Eqs. (\ref{infaff}) and (\ref{fullent}), the affinity related to an information reservoir is   
\begin{equation}
\mathcal{F}_n=\ln[(1-\epsilon_n)/\epsilon_n],
\label{AFF}
\end{equation}
with the associated current being $J_n=-J_{ij}^{(n)}$.
The variable $\xi$ in the formulas below can be either the index $k$ or the index $n$, so that $\sum_\xi= \sum_k+\sum_n$.
Near equilibrium, where all affinities are close to zero, a flux can be written as
\begin{equation}
J_\xi= \sum_{\xi'}L_{\xi\xi'}\mathcal{F}_{\xi'},
\label{linear2}
\end{equation}
where  
\begin{equation}
L_{\xi\xi'}\equiv \left.\frac{\partial J_\xi}{\partial\mathcal{F}_{\xi'}}\right|_{\mathcal{F}=0}
\end{equation}
is the Onsager coefficient, with $\mathcal{F}$ representing a vector with all affinities. The standard entropy production (\ref{fullent}) then becomes
\begin{equation}
\dot{s}= \sum_{\xi\xi'}L_{\xi\xi'}\mathcal{F}_\xi\mathcal{F}_{\xi'}.
\end{equation}

From (\ref{infaff}), (\ref{gammadef}), and (\ref{pijdef}), the current related to reservoir $n$, as given in Eq. (\ref{gencurr}), can be written as
\begin{equation}
J_{n}=-J_{ij}^{(n)}= \gamma_n(p_{n}-\epsilon_n),
\label{currconv}
\end{equation}
which leads to
\begin{equation}
p_n= \epsilon_n+\frac{J_n}{\gamma_n}.
\end{equation} 
Assuming $p_n-\epsilon_n$ small, we expand the rate of Shannon entropy change (\ref{sharate}) in the following way, 
\begin{eqnarray}
\dot{h}_n &=& \mathcal{F}_nJ_n-\frac{J_n^2}{2 \gamma_n\epsilon_n(1-\epsilon_n)}+\textrm{O}(J_n)^3\nonumber\\
&=&\mathcal{F}_nJ_n-2\frac{J_n^2}{\gamma_n}+\textrm{O}(J_n)^3,
\label{linear1}
\end{eqnarray}
where we set $\epsilon_n=1/2$ for the term $\epsilon_n(1-\epsilon_n)$. The choice $\epsilon_n=1/2$ corresponds to the
genuine equilibrium of the system, with the affinity in Eq. (\ref{AFF}) being $\mathcal{F}_n=0$. On the other hand, $\epsilon_n=p_n\neq 1/2$
corresponds to a ``stall force'' case. Hence, setting $\epsilon_n=1/2$ in Eq. (\ref{linear1}) implies  
a linear response treatment with respect to genuine equilibrium.
Using Eqs. (\ref{entinfphys}), (\ref{linear2}), and (\ref{linear1}), we obtain the IP-entropy production 
in the linear response regime,
\begin{equation}
\dot{s}_1= \sum_{\xi\xi'}\left(L_{\xi\xi'}-2\sum_n \frac{L_{n\xi}L_{n\xi'}}{\gamma_n}\right)\mathcal{F}_\xi\mathcal{F}_{\xi'}.
\end{equation}
Note that $\gamma_n= (P_i+P_j)(W_{ij}^{(n)}+W_{ji}^{(n)})$ can be obtained from the equilibrium probabilities with $\mathcal{F}=0$.

\subsection{IP-efficiency at maximum power}

A well known result in linear response theory is that the efficiency at maximum power for uni-cyclic machines is $1/2$ \cite{seif12}.
We now calculate the IP-efficiencies at maximum power for uni-cyclic machines. The standard efficiency contains the
work entering the system in its denominator, which corresponds to the work to reset the tape appearing in the
standard entropy production, as explained in Sec. \ref{sec2}.  Since this work is larger than the rate of Shannon entropy change, the IP-efficiency 
at maximum power should not be smaller than $1/2$. 

\begin{figure}
\includegraphics[width=72mm]{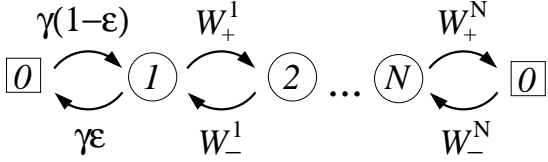}
\vspace{-2mm}
\caption{Uni-cyclic model. The transition rates between state $0$ and $1$ are associated with an information reservoir, whereas the other transition rates
are related to a standard reservoir. Note that we have a cyclic system with $N+1$ being the $0$ state again.  
}
\label{fig8} 
\end{figure}
 
We consider the generic uni-cyclic machine with $N+1$ states on a ring depicted in Fig. \ref{fig8}. The transition rates between states $0$ and $1$, which are related to the information reservoir, are
$\gamma(1-\epsilon)$ and $\gamma\epsilon$, with the first being from $0$ to $1$. The other transition rates are related to standard reservoirs with inverse temperature $\beta=1$, and the 
transition rate from $n$ ($n+1$) to $n+1$ ($n$) is $W_+^n$ ($W_-^n$). We assume that the affinity 
\begin{equation}
\mathcal{F}_{\textrm{out}}= \ln(W_-/W_+),
\end{equation}     
where $W_+= \prod_{n=1}^{N} W_+^n$ and $W_-= \prod_{n=1}^{N} W_-^n$, is related to work extracted from the system.

For the system to operate as a machine the probability current from left to right in Fig. \ref{fig8} must be positive. This probability current is
\begin{equation}
J= \gamma(P_0(1-\epsilon)-P_1\epsilon)= \gamma(P_1+P_0)(p_\tau-\epsilon),
\label{curruni}
\end{equation}
where $p_\tau\equiv P_0/(P_0+P_1)$. The affinity related to the information reservoir is
\begin{equation}
\mathcal{F}_\epsilon= \ln[(1-\epsilon)/\epsilon].
\end{equation}
It is convenient to define
\begin{equation}
C\equiv \sum_{a=0}^{N-1}\left(\prod_{n=1}^{a}W_-^n\right)\prod_{m=a+2}^NW_+^m
\end{equation}
and
\begin{equation}
p\equiv 1/(1+\exp\mathcal{F}_{\textrm{out}}).
\end{equation}
Using a diagrammatic method to obtain the stationary probability distribution \cite{hill}, we obtain  
\begin{equation}
p_\tau= \tau p+(1-\tau)\epsilon 
\end{equation}
where $\tau\equiv k'/(k'+\gamma')$, with $k'\equiv W_-/C$ and $\gamma'\equiv \gamma/[1+\exp(-\mathcal{F}_{\textrm{out}})]$. Note that this formula is similar to the formula (\ref{ptaueq}) for the two 
state model of Sec. \ref{sec2}, which corresponds to $N=1$. The parameter $k'$ has dimension of a transition rate and is related to the thermal transition rates. Therefore, the parameter $0\le \tau\le 1$
is dimensionless being $1$ ($0$) if the transitions of the information reservoir, which are proportional to $\gamma$, are much slower (faster) than thermal transitions.    

Up to first order in the affinities, the current (\ref{curruni}) becomes
\begin{equation}
J= \gamma(P_1+P_0)\tau(p-\epsilon)\approx \Gamma(\mathcal{F}_\epsilon-\mathcal{F}_{\textrm{out}}),
\end{equation}
where $\Gamma\equiv \gamma(P_0+P_1)\tau/4$. Hence, within linear response, the rate of extracted work is
\begin{equation}
\dot{w}_{\textrm{out}}=\mathcal{F}_{\textrm{out}} J= \Gamma(\mathcal{F}_\epsilon-\mathcal{F}_{\textrm{out}})\mathcal{F}_{\textrm{out}},
\label{wweq}
\end{equation}  
and the rate of Shannon entropy change (\ref{linear1}) is
\begin{equation}
\dot{h}= \Gamma(\mathcal{F}_\epsilon-\mathcal{F}_{\textrm{out}})[\mathcal{F}_\epsilon-\frac{\tau}{2}(\mathcal{F}_\epsilon-\mathcal{F}_{\textrm{out}})].
\label{hheq}
\end{equation}  

We now maximize the power $\dot{w}_{\textrm{out}}$ with respect to the output $\mathcal{F}_{\textrm{out}}$ for fixed input $\mathcal{F}_\epsilon$. The power is maximum at
$\mathcal{F}_{\textrm{out}}^*= \mathcal{F}_\epsilon/2$, which gives the IP-efficiency at maximum power 
\begin{equation}
\eta^*\equiv \frac{\dot{w}_{\textrm{out}}^*}{\dot{h}^*}= \frac{1}{2-\tau/2},
\label{maxpower}
\end{equation}
where  $\dot{w}_{\textrm{out}}^*$ and $\dot{h}^*$ are obtained from (\ref{wweq}) and (\ref{hheq}) with $\mathcal{F}_{\textrm{out}}^*= \mathcal{F}_\epsilon/2$, respectively.
The IP-efficiency at maximum power reaches its maximum value $2/3$ for $\tau\to1$, where the transitions related to the information reservoir are much slower than the thermal transitions.
If we had taken the work to reset the tape $\mathcal{F}_\epsilon J$ in the denominator, leading to the usual efficiency based on $\dot{s}$, the standard result $1/2$ would have been obtained. 

\subsection{IP-efficiency at maximum erasure rate}

Another interesting case is the IP-efficiency at maximum erasure rate when the system operates as an eraser, i.e., $J'=-J\ge0$.
The work entering the system to erase the tape is
\begin{equation}
\dot{w}=\mathcal{F}_{\textrm{in}} J'= \Gamma(\mathcal{F}_{\textrm{in}}-\mathcal{F}_\epsilon)\mathcal{F}_{\textrm{in}},
\label{wwweq}
\end{equation}  
where $\mathcal{F}_{\textrm{in}}= \mathcal{F}_{\textrm{out}}$. Rewriting (\ref{hheq}), the erasure rate becomes 
\begin{equation}
-\dot{h}= \Gamma(\mathcal{F}_{\textrm{in}}-\mathcal{F}_\epsilon)[\mathcal{F}_\epsilon+\frac{\tau}{2}(\mathcal{F}_{\textrm{in}}-\mathcal{F}_\epsilon)].
\label{hhheq}
\end{equation}   
Maximizing the erasure rate with respect to $\mathcal{F}_\epsilon$ for fixed input, we obtain that $-\dot{h}$ is maximal at
$\mathcal{F}_{\epsilon}^\dagger= \mathcal{F}_{\textrm{in}}(1-\tau)/(2-\tau)$. The IP-efficiency at maximum erasure rate is then
\begin{equation}
\eta^\dagger\equiv \frac{-\dot{h}^\dagger}{\dot{w}^\dagger}= \frac{1}{2},
\label{maxer}
\end{equation}
where  $\dot{w}^\dagger$ and $-\dot{h}^\dagger$ are evaluated at $\mathcal{F}_\epsilon= \mathcal{F}_\epsilon^\dagger$. Note that this efficiency, unlike (\ref{maxpower})   
is independent of $\tau$ whereas $\mathcal{F}_{\epsilon}^\dagger= \mathcal{F}_{\textrm{in}}(1-\tau)/(2-\tau)$, unlike $\mathcal{F}_{\textrm{out}}^*= \mathcal{F}_\epsilon/2$,
depends on $\tau$. In \cite{bara13} we have obtained an efficiency at maximum erasure for a specific model of a system interacting with a tape that could move in both directions. The result obtained
in this reference was $1/3$. The difference with the present result comes from the fact that in \cite{bara13} we have considered an extra term in the denominator which was
related to the possibility of taking back a bit from the outgoing tape to interact with the system.

\section{Conclusion}
\label{sec6}

We have generalized the theory of stochastic thermodynamics to include information reservoirs. Such reservoirs can be understood as a tape that has its Shannon entropy modified due
to the interaction with the system but does not exchange energy with the system. Thus information reservoirs contribute to the second law while leaving the first law unaltered. This generalization
is achieved with the IP-entropy production, which differs from the standard entropy production of stochastic thermodynamics. Both entropy productions follow from the more general
inequality (\ref{gensecond}), which can be further generalized with the fluctuation theorem proved in App. \ref{appa}. 

In principle, with our framework any thermodynamic system interacting with information reservoirs can be studied. Our theory allows for the construction of simple models that can be used to understand the
qualitative behavior of a thermodynamic system interacting with an information reservoir. For example, with the three-state model for a thermoelectric effect of Sec. \ref{sec4}, we have shown that 
that there are regions in the phase diagram where the system can take heat from the cold reservoir and drive particles against the chemical potential gradient. Furthermore, a convenient feature
is that the full thermodynamic cost to reset the tapes to their original configurations is easily accessible, being contained in the standard entropy production.   

The power of our approach is also demonstrated by the fact that it allowed for the development of a systematic linear response theory for information processing machines, 
which was still lacking in the literature. As main results, we have obtained the IP-entropy production in the linear response regime in terms of
the Onsager coefficients and the affinities, and we have obtained IP-efficiencies (at maximum power and maximum erasure rate) for uni-cyclic machines.

\begin{acknowledgments}
We thank D. Hartich for helpful discussions. 
\end{acknowledgments}

\appendix

\section{Fluctuation theorem generalizing inequality (\ref{gensecond})}
\label{appa}

We prove a fluctuation theorem leading to the inequality (\ref{gensecond}). We consider  a generic Markov jump process with transition rates denoted by $W_{ij}$. The number of states is
duplicated, with state $i$ being duplicated to $i_A$ and $i_B$. The transition rates in the new duplicated system are such that states with the same subscript are not connected, i.e., the transition rates 
between them are zero. The transition rates in the duplicated system are related to the transition rates in the original system system by the formula $W_{i_Aj_B}=W_{i_Bj_A}= W_{ij}$. Moreover, the transition
rates between $i_A$ and $i_B$ are $W_{i_Ai_B}= W_{i_Bi_A}= R_i$. 

The stationary probability in the duplicated system is the same as in the original system. More precisely, the stationary master equation for $P_{i_A}$ in the duplicated system is
\begin{equation}
\sum_{j\neq i}\left(P_{j_B}W_{ji}-P_{i_A}W_{ij}\right)+\left(P_{i_B}-P_{i_A}\right)R_i=0.
\end{equation}  
Comparing with Eq. (\ref{meq}), we see that $P_i=P_{i_A}+P_{i_B}$, where $P_i$ indicates the stationary probability of state $i$ in the original system. A definition that is
useful for the discussion below is the escape rate of state $i_A$
\begin{equation}
\lambda(i_A)\equiv \sum_{j\neq i}W_{ij}+R_i.
\label{escaprat}
\end{equation} 
Note that $\lambda(i_A)= \lambda(i_B)$.

A stochastic trajectory in the duplicated system for a time interval $t\in[0,T]$ is denoted $X_T=(x_0,\tau_0;x_1,\tau_1;\ldots;x_N,\tau_N)$, where $x_n$ is the state for $t\in[t_{n},t_{n}+\tau_n]$, with $t_0=0$, $t_{n+1}= t_n+\tau_n$,
and $t_{N+1}= T$. The probability of a trajectory is
\begin{equation}
\mathcal{P}[X_T]= P(x_0)\left(\prod_{n=0}^{N-1}W_{x_nx_{n+1}}\right)\prod_{n=0}^{N}\exp(-\lambda(x_n)\tau_n)
\end{equation} 
where $P(x_0)$ denotes the initial probability. The probability of the reversed trajectory $\tilde{X}_T=(x_N,\tau_N;\ldots;x_1,\tau_1;x_0,\tau_0)$ with modified 
transition rates $\overline{W}_{ij}$ (or $\overline{R}_i$ is the jump is between $i_A$ and $i_B$)  reads       
\begin{equation}
\overline{\mathcal{P}}[\tilde{X}_T]= \tilde{P}(x_N)\left(\prod_{n=0}^{N-1}\overline{W}_{x_{n+1}x_n}\right)\prod_{n=0}^{N}\exp(-\overline{\lambda}(x_n)\tau_n),
\end{equation} 
where $\tilde{P}(x_N)$ is the initial probability of the reversed trajectory and $\overline{\lambda}(x_n)$ is the escape rate for the modified rates.
The ratio of trajectory probabilities then becomes 
\begin {eqnarray}
\frac{\mathcal{P}[X_T]}{\overline{\mathcal{P}}[\tilde{X}_T]}= & \frac{P(x_0)}{\tilde{P}(x_N)}\left(\prod_{n=0}^{N-1}\frac{W_{x_nx_{n+1}}}{\overline{W}_{x_{n+1}x_n}}\right)\nonumber\\
&\times\prod_{n=0}^{N}\exp[(\overline{\lambda}(x_n)-\lambda(x_n))\tau_n)].
\label{ratioeq}
\end{eqnarray}
From Eq. (\ref{escaprat}), we obtain that the term $\overline{\lambda}(x_n)-\lambda(x_n)=0$ if
\begin{equation}
R_{i}+\sum_{j\neq i}W_{ij}=\overline{R}_{i}+\sum_{j\neq i}\overline{W}_{ij},
\label{Wconstraint2}
\end{equation}
which is the constraint (\ref{Wconstraint}). 

The activity for jumps from $i_A$ to $j_B$ and from $i_B$ to $j_A$ is a functional of the the trajectory $X_T$ defined as
\begin{equation}
\mathcal{K}_{ij}[X_T]\equiv \sum_{n=0}^{N}\left(\delta_{x_n,i_A}\delta_{x_{n+1},j_B}+\delta_{x_n,i_B}\delta_{x_{n+1},j_A}\right).
\end{equation}
With this activity we define the functional  
\begin{equation}
\Omega[X_T]\equiv \sum_i\left(\sum_{j\neq i}\mathcal{K}_{ij}[X_T]\ln\frac{W_{ij}}{\overline{W}_{ji}}+\mathcal{K}_{ii}[X_T]\ln\frac{R_i}{\overline{R}_i}\right).
\end{equation}
If the constraint (\ref{Wconstraint2}) is satisfied, by choosing uniform distributions for both $P(x_0)$ and $\tilde{P}(x_N)$ in Eq. (\ref{ratioeq}), we obtain
\begin{equation}
\frac{\mathcal{P}[X_T]}{\overline{\mathcal{P}}[\tilde{X}_T]}= \exp\left(\Omega[X_T]\right).
\end{equation}
This relation then implies 
\begin{equation}
\langle\exp(-\Omega)\rangle\equiv \sum_{X_T}\exp\left(-\Omega[X_T]\right)\mathcal{P}[X_T]=\sum_{X_T}\overline{\mathcal{P}}[\tilde{X}_T]=1,
\end{equation}  
where $\sum_{X_T}$ represents an integral over all stochastic trajectories. This integral fluctuation theorem leads to the inequality
\begin{equation}
\langle\Omega\rangle/T\ge 0.
\end{equation}
The above inequality is equivalent to (\ref{gensecond}), as $\langle\Omega\rangle/T= \dot{\omega}$. We note that the functional $\Omega$ is, in general, not 
antisymmetric, i.e., it cannot be written as a sum of probability currents. It does become antisymmetric if the auxiliary rates are chosen so that $\langle\Omega\rangle/T$
becomes the standard entropy production but for auxiliary rates leading to the IP-entropy production it does not. 

Whereas the standard entropy production $\dot{s}$ can be obtained from a fluctuation theorem for the original system \cite{seif12}, in order to obtain the IP-entropy production $\dot{s}_1$ 
we need this fluctuation theorem for the duplicated system. This duplication has a physical interpretation if we compare Fig. \ref{fig3} with Fig. \ref{fig4} for the paradigmatic model of Sec. \ref{sec2}.
The duplication in Fig. \ref{fig3} is necessary to include the possibility of transitions from $u_A$ to $u_B$ and $d_A$ to $d_B$, which corresponds to transitions 
where the new incoming bit is in a state that couples to the state of the system. Note that in the duplicated system of Fig \ref{fig3} the states in the same replica are connected by the thermal transition link, which is different from the 
duplication in this appendix. If a set of links is assumed to be related to standard reservoirs, then a duplicated system keeping these links connecting states in the same replica and not in different replicas suffices to obtain
a fluctuation theorem leading to the corresponding $\dot{s}_1$ \cite{bara14}. The derivation of the fluctuation theorem in this case is very similar. 
Here, we have chosen the duplication scheme above in order to obtain the most general inequality.


\end{document}